      [1999/12/01 v1.4c Il Nuovo Cimento]
\documentclass{cimento}

\usepackage{graphicx}  
\title{Tracing the distribution and evolution of Iron in the 
IntraCluster Medium}
\author{P. ~ Tozzi\from{ins:x}}
\instlist{\inst{ins:x}INAF, Osservatorio Astronomico di Trieste}
\PACSes{\PACSit{98.65.Cw}
\PACSit{98.65.Cw}{\ldots}}
\begin{document}

\maketitle

\begin{abstract}
Emission lines in X--ray spectra of clusters of galaxies reveal the
presence of heavy elements in the diffuse hot plasma (the Intra
Cluster Medium, or ICM) in virial equilibrium in the dark matter
potential well.  Thanks to the X--ray satellites {\sl Chandra} and
{\sl XMM--Newton} we are now able to measure with good accuracy the
distribution and evolution of Iron up to redshift $z\sim 1.3$.  The
capability of studying the chemical and thermodynamical properties of
the ICM in high redshift clusters is an efficient tool to constrain
the interaction processes between the cluster galaxies and the
surrounding medium.  We confirm that the ICM is already significantly
enriched at a look-back time of 9~Gyr, and find that the Iron
abundance change with redshift as $\propto (1+z)^{-1.25}$, implying an
increase of a factor of $\sim 2$ with respect to $z\simeq 1.3$.  This
result can be explained by a prompt enrichment by star formation
processes in massive ellipticals at $z\geq 2$, followed by a slower
release of enriched gas from disk galaxies into the ICM, associated to
a morphological transition from disk to S0.
\end{abstract}

\section{X--ray Clusters of Galaxies as laboratories for
galaxy evolution and signposts for cosmology}

Clusters of galaxies are the largest virialized objects in the
Universe.  They form via gravitational instability from the initial
perturbations in the matter density field.  Most of their baryonic
content is in the form of diffuse gas.  The diffuse baryons are heated
to temperatures between $kT \sim 1$ keV and $kT \sim 10$ keV (roughly
corresponding to masses ranging from $10^{14}$ to $5 \times 10^{15}
M_\odot$) and constitute the Intra Cluster Medium (from now on ICM).
Thanks to the very low density of the electrons (typically $n_e
\sim 10^{-3}$ cm$^{-3}$) the ICM is optically thin and it is in a
state of collisional equilibrium established between the electrons and
the heavy ions.  The resulting emission from the ICM in the X--ray
band is described by a continuum component due to thermal
Bremsstrahlung (roughly scaling as $\propto T^{1/2} n_e^2$) plus line
emission from K--shell and L--shell transitions of heavy ions (Iron
being the most prominent).  The collisional equilibrium allows one to
derive directly from the X--ray spectrum of the ICM both the electron
temperature and the abundances of heavy elements.  

\begin{figure}[t!]
\centering \includegraphics[height=6cm]{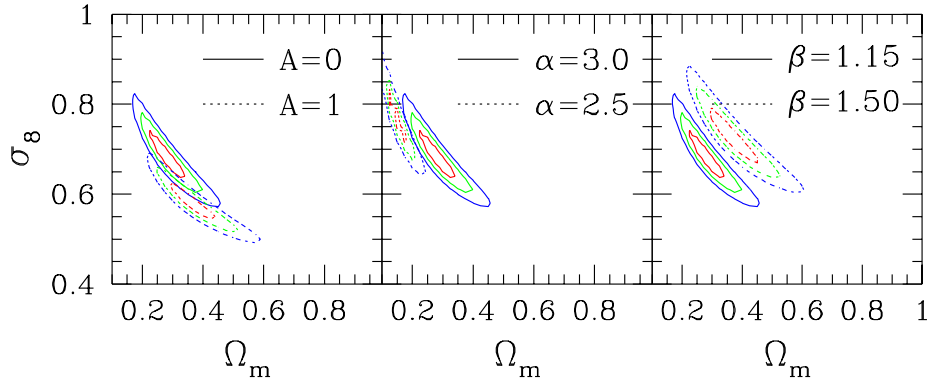}
\caption{\footnotesize Confidence levels in the $\sigma_8$--$\Omega_m$
plane from the X--ray luminosity function of distant clusters of
galaxies observed with ROSAT.  Different panels correspond to
different parametrization of the $L$--$T$ relation (see
\cite{bor99}).}
\label{sigma8}
\end{figure}

X--ray cluster samples have always been considered one of the best
tools for cosmology, thanks to the simple link between the X--ray
spectral properties (namely the electron temperature) and the total
dynamical mass, and to the simple selection function depending only on
the flux limit.  Their number density as a function of the mass scale
and of the cosmic epoch, strongly depend on cosmological parameters.
The number density of massive (hot) clusters is expected to decrease
at high redshift at a rate sensibly depending on cosmological
parameters (see \cite{r02}).  In this respect, the detection of high
redshift clusters has a very strong impact.  About fifteen years ago,
before results from deep surveys with the ROSAT satellite had been
published \cite{R98}, the presence of massive, hot clusters at
redshift as low as $z\sim 0.2$ was strongly questioned.  Now, the
presence of a significant number of massive clusters at redshift as
large as $z\sim 1.4$ is firmly established, and this represents a
strong hint in favour of a low density, $\Omega_0\sim 0.3$, $\Lambda
\sim 0.7$ CDM Universe.  The highest-z cluster selected so far in the
X--ray band is at redshift $z\sim 1.4$ \cite{stan06}.  In particular,
we would like to remark that the value of the normalization of the
initial density perturbation field $\sigma_8\sim 0.7$ measured with
X--ray clusters (\cite{bor99}, see Figure \ref{sigma8}), initially at
variance with the measure $\sigma_8 = 0.84 \pm 0.04$ in the first year
of WMAP, is now in agreement with the new WMAP results \cite{spe07}.
Once again, this is a good example of the complementarity of X--ray
clusters with respect to other geometrical cosmological tests, like
high--z SNe and CMB experiments.

Now that we are in the maturity of the {\sl Chandra} and {\sl
XMM--Newton} era, we realize that clusters harbor an unexpected
complexity, that demands a much deeper understanding of the
thermodynamics of the ICM and its relation with the other mass
components, such as member galaxies and dark matter. One of the
biggest evidence of this complexity is given by the cool core problem.
In more than half of local clusters the cooling time within 10 kpc can
be significantly less than 1 Gyr \cite{pet06}.  It seems unavoidable
to predict that baryons are flowing to the cold phase at a rate of the
order of 100, sometimes 1000 $M_\odot$ yr$^{-1}$.  The simplest model
based on isobaric cooling, predicts a spectrum rich in emission lines,
which are strongly increasing at low temperatures due to the higher
number of ion species.  However, grating spectroscopy of the brightest
central regions of clusters, with the {\sl XMM--Newton} satellite,
provided a surprising result.  Many of the lines expected in cooling
flows were missing from the observed spectra \cite{pet01}.  It can be
shown that the lowest temperature in the center is of the order of 1/3
of the virial one.

This discovery has a strong impact: it implies that the ICM is kept
above a temperature floor, not too far from the virial one, by some
heating mechanism.  Which is the process that keep the temperature of
the ICM above $1/3 \, T_{vir}$?  The presence of some extra--energy in
the ICM was already known from the study of the scaling relations
between X--ray observables and from the observation of an entropy
excess in the ICM with respect to the self--similar scaling
\cite{pon99}.  However, there is no consensus on the sources which
constantly heat the ICM.  This is an open problem, relevant not only
to the ICM, but also for general framework of galaxy formation and
evolution.  The main problem with feedback, is that any process we may
think of, scales with volume (and then with density $\rho$), while
cooling is a runaway process proportional to $\rho^2$.  Therefore
there is not an obvious, mechanism for self--regulation.
Understanding feedback is nowaday the most compelling goal for
structure formation models.  Main candidates are SNe explosions and
stellar winds (as confirmed by the presence of heavy elements in the
ICM) and the much more energetic output from AGN.  A spectacular
example that favours AGN as the main heating sources, is the X--ray
image of the Persues cluster \cite{fab06}, where hot bubbles created by
the jets are pushing the ICM apart, with a total mechanical energy
sufficient to heat significantly the diffuse baryons. Still, how and
on which time scale the energy is thermalized into the ICM is still a
matter of debate.

It is worth mentioning that the presence of the non--gravitational
energy input does not hamper the use of clusters as a tool for
cosmology.  The relation between X--ray observables and the dynamical
mass is still valid, and maybe it is even tighter once we provide a
comprehensive model for the ICM thermodynamics.  A recent example is
the use of the parameter $T_X M_{gas}$ which appears to be tightly
correlated to the dynamical mass \cite{kr06}.  Therefore, this
increasing complexity is not an impediment, but, on the contrary, is
giving us a more powerful tool to investigate at the same time many
aspects of structure formation, from subgalactic to cluster scales.
The price to pay is, of course, a bigger effort in describing the ICM
physics.

In this Contribution, I will show how the study of the evolution of
Iron abundance with redshift through detailed X--ray spectroscopy of
the ICM may help to shed light on the interaction processes between
cluster galaxies and the ICM itself.

\section{Tracing the evolution of Iron abundance in the ICM}

The presence of heavy elements in the ICM traces the distribution of
SNe products into the diffuse hot baryons.  Therefore, the evolution
of heavy elements abundances with cosmic epoch provides a useful
fossil record for the past star formation history of cluster baryons
and constraints on the interaction of the ICM with cluster galaxies.
So far, only the abundance of Iron can be traced back to the highest
redshifts where X--ray clusters are observed.  Over a wide range of
temperatures, the Equivalent Width (EW) of the Fe K--shell line
(mostly from Fe + 24 and Fe +25) is several orders of magnitudes
larger than any other spectral feature.  At lower energies O, Si, S,
and L--shell transition in lower Fe ions show significant EW.  Their
abundance is consistent with being produced by the elliptical cluster
galaxies \cite{mat88}.  The ratio of the abundance of the $\alpha$
elements over Fe is very relevant to understand the relative
contribution of TypeII and TypeIa SNe.  However, here we will discuss
only some results on the Fe abundance at high redshift.

\begin{figure}
\centering
\includegraphics[width=6 cm, angle=0]{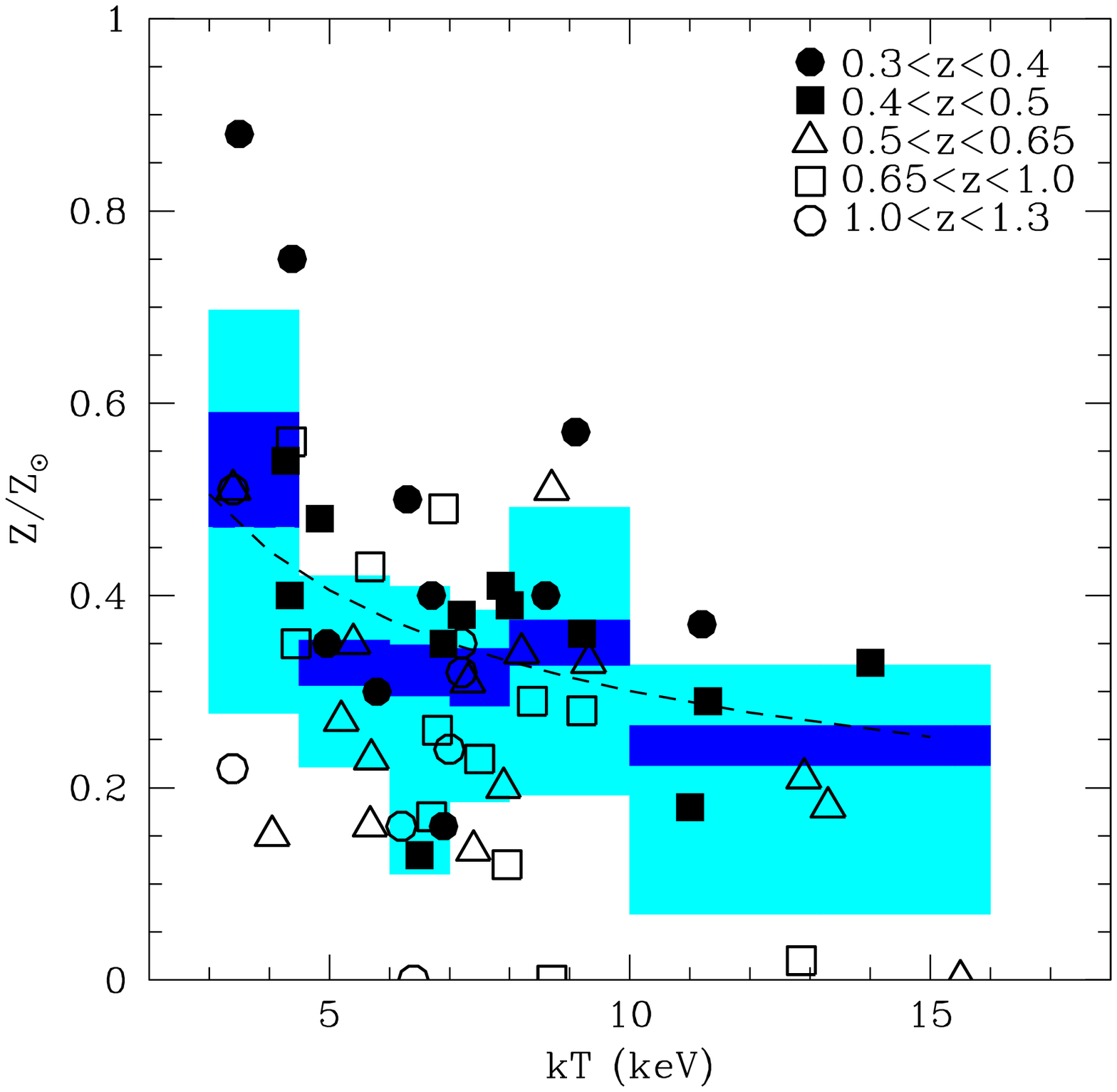}
\includegraphics[width=6 cm, angle=0]{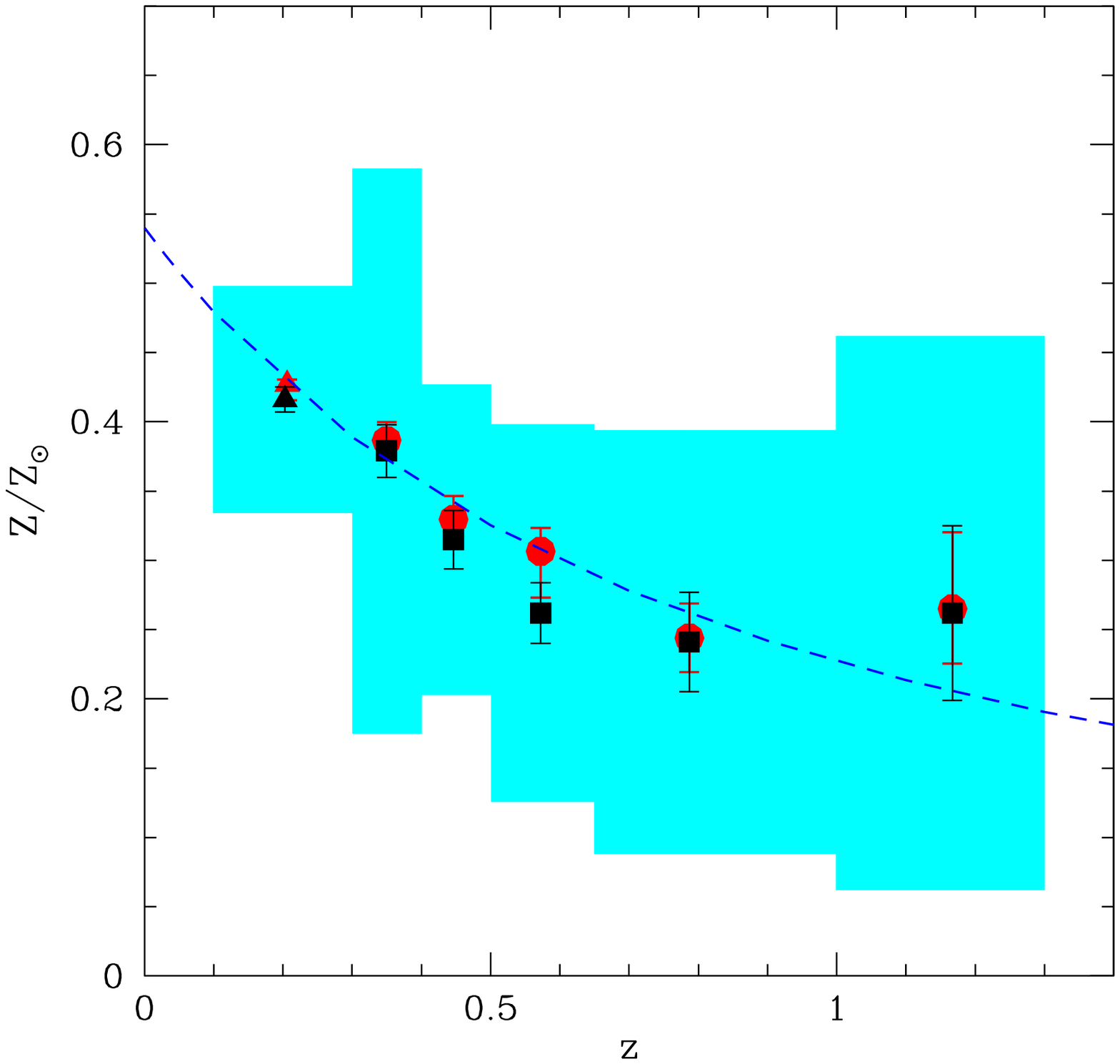}
\caption{{\em Left:} scatter plot of best-fit $Z_{Fe}$ values 
versus $kT$. The dashed line represents the best-fit $Z_{Fe}-T$
relation ($Z_{Fe}/Z_\odot\simeq0.88\,T^{-0.47}$). Shaded areas show
the weighted mean (blue) and average $Z_{Fe}$ with {\em rms}
dispersion (cyan) in 6 temperature bins.  {\em Right:} mean $Z_{Fe}$
from combined fits (red circles) and weighted average of single-source
measurements (black squares) within 6 redshift bins. The triangles at
$z\simeq0.2$ are based on the low-z sample described in
\cite{b07}. Error bars refer to the $1\sigma$ c.l.. Shaded areas
show the {\em rms} dispersion.  The dashed line indicates the best fit
over the 6 redshift bins for a simple power law of the form $\langle
Z_{Fe}\rangle=Z_{Fe}(0)\,(1+z)^{-1.25}$ \cite{bal07}.  Solar
abundances refer to \cite{angr89}.}
\label{fig02}
\end{figure}

In a recent work \cite{bal07} we used the {\em Chandra} archive for
clusters at redshift $z\geq 0.4$ to compute the average Fe abundance
in several redshift bins.  Our analysis suggests higher Iron
abundances at lower temperatures in all the redshift bins. This trend
is somewhat blurred by the large scatter.  The correlation is more
evident when we compute the weighted average of the Iron abundance
$Z_{Fe}$ in 6 temperature intervals, as shown by the shaded areas in
Figure ~\ref{fig02} (left panel).

The $Z_{Fe}$ measured from the combined fits in 6 redshift bins is
also shown in Figure ~\ref{fig02} (right panel). We computed the
weighted average from the single cluster fits in the same redshift
bins.  The best-fit values resulting from the combined fits are always
consistent with the weighted means within $1\sigma$.  The average Iron
abundance is already significant at $z\sim 1.3$, at a look-back time
of $\sim 9 $ Gyr.  In addition, we observe a significant increase of
the average Iron abundance with cosmic time below $z=0.5$.  The
evolution in $Z_{Fe}$ with $z$ can be parametrized by a power law of
the form $\sim(1+z)^{-1.25}$.  The observed evolution implies that the
average Iron content in the inner regions (radii $R \leq 0.2 R_{vir}$)
of the ICM at the present epoch is a factor of $\sim2$ larger than at
$z\simeq 1.3$.

\begin{figure}
\centering
\includegraphics[width=7 cm, angle=0]{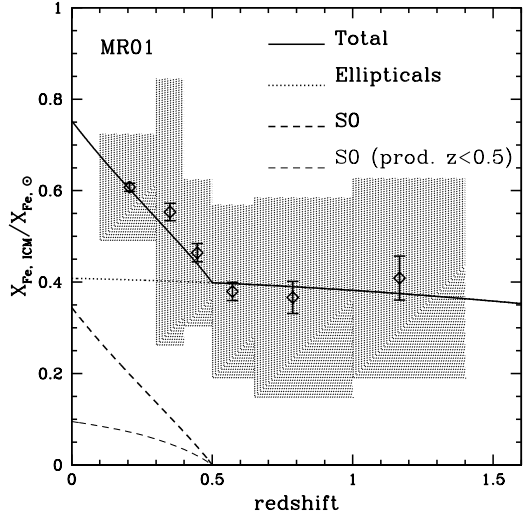}
\caption{Observed and predicted redshift evolution of the Iron abundance 
in the ICM relative to solar.  The open diamonds with error bars and
the shaded areas show the weighted means and the \emph{rms} dispersion
around the weighted means of the Iron abundances in the 5 redshift
bins as observed by \cite{bal07}, respectively.  The Iron abundances
have been rescaled by a factor of 1.5 with respect to figure
\ref{fig02}, to be consistent with the recent solar values by
\cite{aspl05}.  The thick dashed lines, dotted lines and the solid
lines are the predicted contributions to the ICM Iron abundance given
by S0 galaxies, ellipticals and the total ICM Iron abundance,
respectively, calculated assuming for the Ia SN rate by
\cite{MR01}.  The thin dashed lines represent the contribution due to
the Iron produced in S0 at $z<0.5$ \cite{cal07}.}
\label{calura}
\end{figure}

This simple result requires a complex interpretation.  On one hand, we
notice that the Fe abundance is already significant at $z>1$, in line
with the expectation that the bulk of star formation in massive
spheroids, responsible for the large majority of the metals, occurs at
$z \geq 2$.  On the other hand, we do not expect much star formation
responsible for Iron production after $z\sim 0.5$ (but see
\cite{ett05}).  Therefore, the most likely interpretation of the
increase of the average Iron abundance in the inner $0.2 R_{vir}$ of
clusters, may be due to deposition of previously enriched gas towards
the center.  Currently, several approaches can be used.  A
phenomenological model based on detailed chemical galactic evolution,
shows that the increase of Iron is consistent with the transformation
of gas--rich disk galaxies into S0 galaxies (see Figure \ref{calura}
and \cite{cal07}), with the consequent deposition of highly enriched
gas in the central regions where the ram pressure stripping and other
dynamical mechanisms are more efficient in removing gas from the
disks.  N--body Hydrodynamic simulations, on the other hand, show that
high abundance, low entropy gas, previously associated to galaxies or
group--size halos, may sink to the center during the mass growth of
the cluster (Cora et al. in preparation).  These models favour a
dynamical origin of the observed evolution, but they must also explain
the abundance profiles and the temperature gradients in cluster cores
observed in spatially resolved spectral analysis.  In particular, the
presence of cool cores is always associated to strong peak in Iron
abundance \cite{deg04} and the slope of the inner temperature gradient
strongly correlates with the ratio between the cooling time and the
age of the universe at the cluster redshift \cite{b07}.  For this
reason, in \cite{bal07} we also investigated whether the observed
evolution is driven by a negative evolution in the occurrence of
cool-core clusters with strong metallicity gradients, but we do not
find any clear evidence of this effect in the data.  On the other
hand, detailed studies on the surface brightness of distant X--ray
clusters, show that moderate cool core are already present at $z>1$
(Santos et al. in preparation).  Overall, these results provide
significant constraints on the time scales of the physical processes
that drive the chemical enrichment of the ICM.  However, the mechanism
responsible for the gradual transfer of the enriched gas from the
cold, low entropy phase associated to galaxy or group--size halos to
the ICM is still under investigation.

\section{Conclusions}

Precise measurements of the Iron content of clusters over large
look--back times provide a useful fossil record for the past star
formation history of cluster baryons.  In a recent work \cite{bal07}
we confirm the presence of a significant amount of Iron in high-$z$
clusters.  We find significant evidence of a decrease in $Z_{Fe}$ as a
function of redshift, which can be parametrized by a power law
$\langle Z_{Fe}\rangle \simeq (0.54\pm 0.04)Z_\odot \,(1+z)^{-1.25}$
(where solar abundances refer to \cite{angr89}).
This result can be explained by a prompt enrichment by star formation
in massive ellipticals at $z\geq 2$, followed by a slower
release of enriched gas from disk galaxies, associated to
a morphological transition from disk to S0.  To investigate further
the interaction processes between ICM and cluster galaxies, we need to
understand at the same time the distribution and the evolution of
heavy elements in the ICM, and therefore we rely on high--resolution,
spatially resolved X--ray spectroscopy.

Next future X--ray clusters studies must face a double challenge:
first, understanding the physics of the ICM, exploring in greater
details the cool--cores and the low surface brightness regions in the
outskirts of clusters; second, finding more clusters and groups at
high redshifts.  
To capitalize and extend what we have learned so far, we must have
soon both a wide area medium--deep survey as well as a mission devoted
to the properties of the ICM.  Looking further into the future, the
next generation of large X--ray telescopes must achieve a spatial
resolution comparable to that of {\sl Chandra}, which proved to be
crucial to avoid confusion limit and to isolate genuine diffuse
emission at very high redshifts.

\acknowledgments

We would like to thank the organizers of this conference for providing
a stimulating scientific environment.  We acknowledge financial
contribution from contract ASI--INAF I/023/05/0 and from the PD51 INFN
grant.

\end{document}